\documentclass[letterpaper, 10 pt, conference]{ieeeconf}  % Comment this line out if you need a4paper

\IEEEoverridecommandlockouts                              % This command is only needed if 
\usepackage{graphics} % for pdf, bitmapped graphics files
\usepackage{epsfig} % for postscript graphics files
\usepackage{mathptmx} % assumes new font selection scheme installed
\usepackage{times} % assumes new font selection scheme installed
\usepackage{amsmath} % assumes amsmath package installed
\usepackage{amssymb}  % assumes amsmath package installed
\usepackage{mathtools}

\usepackage{caption}
\usepackage{subcaption}

\title{\LARGE \bf
Small Satellite Constellation Separation using Linear Programming based Differential Drag Commands
}

\author{Emmanuel Sin$^{1}$, Murat Arcak$^{2}$ and Andrew Packard$^{3}$% <-this % stops a space
\thanks{$^{1}$Emmanuel Sin is a Graduate Student of the Department of Mechanical Engineering at the University of California, Berkeley
        {\tt\small emansin@berkeley.edu}}%
\thanks{$^{2}$Murat Arcak is a Professor of the Department of Electrical Engineering and Computer Sciences at the University of California, Berkeley
        {\tt\small arcak@berkeley.edu}}%
\thanks{$^{3}$Andrew Packard is a Professor of the Department of Mechanical Engineering at the University of California, Berkeley 
	{\tt\small apackard@berkeley.edu}}%
}        

\begin{document}

\maketitle
\thispagestyle{empty}
\pagestyle{empty}

%%%%%%%%%%%%%%%%%%%%%%%%%%%%%%%%%%%%%%%%%%%%%%%%%%%%%%%%%%%%%%%%%%%%%%%%%%%%%%%%
\begin{abstract}

We study the optimal control of an arbitrarily large constellation of small satellites operating in low Earth orbit. Simulating the lack of on-board propulsion, we limit our actuation to the use of differential drag maneuvers to make in-plane changes to the satellite orbits. We propose an efficient method to separate a cluster of satellites into a desired constellation shape while respecting actuation constraints and maximizing the operational lifetime of the constellation. By posing the problem as a linear program, we solve for the optimal drag commands for each of the satellites on a daily basis with a shrinking-horizon model predictive control approach. We then apply this control strategy in a nonlinear orbital dynamics simulation with a simple, varying atmospheric density model. We demonstrate the ability to control a cluster of 100+ satellites starting at the same initial conditions in a circular low Earth orbit to form an equally spaced constellation (with a relative angular separation error tolerance of one-tenth a degree). The constellation separation task can be executed in 71 days, a time frame that is competitive for the state-of-the-practice. This method allows us to trade the time required to converge to the desired constellation with a sacrifice in the overall constellation lifetime, measured as the maximum altitude loss experienced by one of the satellites in the group after the separation maneuvers. 
\end{abstract}

%%%%%%%%%%%%%%%%%%%%%%%%%%%%%%%%%%%%%%%%%%%%%%%%%%%%%%%%%%%%%%%%%%%%%%%%%%%%%%%%
\section{INTRODUCTION}

Cubesats are miniature satellites comprised of one or more $10\times10\times10$ cm cubic units. Due to the use of commercial off-the-shelf components and a widely accepted reference design, cubesats have become a standard platform for research \cite{CubesatStandard}. As the cost of both satellite manufacturing and launch services decrease, new commercial and scientific applications of cubesats will emerge \cite{CubesatOpps}. In particular, new opportunities may arise from the use of cubesats in large-scale, coordinated constellations \cite{CubesatConstellations}. Coordinated groups of small satellites can enable mission objectives that may be difficult or impossible with single, monolithic satellites (e.g., high-cadence or multipoint measurements, communication relays). In fact, small satellite constellations have already been launched for commercial purposes, such as Earth imaging \cite{Imaging}, and plans have been proposed for their use in providing space-based Internet \cite{Internet1}-\cite{Internet2}.

However, as the number of satellites increases, it becomes more difficult to operate the constellation. Satellite constellation maneuvers can be divided into three broad categories: 1) initial acquisition of desired formation, 2) station-keeping of desired orbital positions and motion in the presence of disturbances, and 3) reconfiguration to a different desired formation. The current state-of-practice requires the control of each individual spacecraft from a command center on the ground that monitors the motion of each satellite. Thus, there exists the need for a centralized and optimal method of controlling satellite constellations from the ground.

Cubesats typically employ a limited actuator suite compared to their larger cousins and may lack propulsive thrusters or other actuators, resulting in limited control authority. In such cases, a cubesat may have to rely on ``passive" means to make orbit maneuvers. One well-studied method called differential drag \cite{DiffDrag1}-\cite{DiffDrag2} employs gyroscopic actuators, such as reaction wheels, to not only change a satellite's orientation but also its orbital in-plane motion, such as its altitude and angular speed. By changing the satellite's cross-sectional area exposed to the incident air molecules in the atmosphere of low Earth orbit, a varying drag force can be applied in the opposite direction of the satellite's orbital velocity. In what is commonly referred to as the drag paradox \cite{DragParadox1}-\cite{DragParadox2}, this drag force makes the satellite fall in altitude but also increase in  angular speed. By causing a differential in the angular speeds of satellites that are in the same orbit, their relative positions can be changed. However, differential drag maneuvers that increase a satellite's velocity also result in a proportional loss in its altitude, effectively reducing its time in orbit. Hence, these differential drag inputs must be applied sparingly so as to not unnecessarily reduce the lifetime of the satellites.

Planet is a San Francisco based Earth imaging company that has successfully used differential drag to form a ``line-scanner" constellation using 3U ($30\times10\times10$) cubesats. The constellation allows Planet to produce a complete image of the Earth's surface every day. Planet describes the use of a bang-bang control approach that commands each satellite to enter either a high-drag or a low-drag window at a certain time and for a specific duration \cite{Planet1}-\cite{Planet2}. This control strategy is considered to be operationally simpler compared to other approaches where the optimal drag area commands may take on any value within the continuous range between low-drag and high-drag. However, as the performance of commercially-available ADCS (attitude determination and control system) sensors and actuators for cubesats increase, it may be of value to investigate different control strategies.

The subsequent sections are organized as follows: Section II begins by introducing the dynamical models used in this paper. In Section III we formulate an optimization problem that we then transform into a linear program to solve for the optimal inputs. Section IV shows our results of applying the linear program on the nonlinear orbital dynamics. We end with Section V where we state concluding remarks.

\section{DYNAMICAL MODELS}

In this section, we first describe the nonlinear ``truth" model that we use for simulating satellite orbital dynamics. We then introduce the approximate discrete-time model to be used in our linear program.

\subsection{Orbital Dynamical Model}

To determine the motion of a satellite orbiting the Earth, we start with the two-body problem where we assume that the barycenter of the system is co-located with the center of a spherically, symmetric Earth (i.e., the mass of the satellite is negligible). The satellite's motion can be described by the following second-order ordinary differential equation \cite{BMW}:
\begin{equation}
 \ddot{\vec{r}} = -\frac{\mu_{\scriptscriptstyle E}}{|\vec{r}|^3}\vec{r} + \vec{a}_{perturb} \\
\end{equation}
where in the first term, $\vec{r}$ is the position vector pointing from the center of the Earth to the satellite and $\mu_{\scriptscriptstyle E}$ is the gravitational parameter of the Earth (i.e., gravitational constant, $G$, multiplied by the mass of the Earth).

While the first term represents the gravitational force exerted on the satellite by the Earth, the second term, $\vec{a}_{perturb}$, represents the specific forces due to perturbations. Examples of such perturbations include the gravitational effects caused by the oblateness of the Earth, gravity from third bodies (e.g., Moon, Sun, other planets), solar radiation pressure and atmospheric drag. 

\subsection{Atmospheric Drag Perturbation Model}

We now describe the atmospheric drag force model from which we derive our control authority. As a satellite moves along its orbit, it experiences an atmospheric drag force that acts against its velocity relative to the atmosphere. The mass-specific acceleration due to this atmospheric drag force is given by the equation \cite{Montenbruck}:
\begin{equation} 
 \vec{a}_{atmdrag}  = -\frac{1}{2}\frac{C_DA}{m}\rho \lvert \vec{v}_{rel} \rvert \cdot \vec{v}_{rel} \\
\end{equation}
where:

\begin{itemize}
\item[] {\makebox[0.4cm]{$C_D$\hfill} : satellite drag coefficient}
\item[] {\makebox[0.4cm]{$A$\hfill} : surface area exposed to incident stream}
\item[] {\makebox[0.4cm]{$m$\hfill} : satellite mass}
\item[] {\makebox[0.4cm]{$\rho$\hfill} : atmospheric density at the satellite position}
\item[] {\makebox[0.4cm]{$\vec{v}_{rel}$\hfill} : velocity of satellite relative to the atmosphere} \\
\end{itemize} 

For the simulations in this paper, we use the same constant values used by Li and Mason \cite{Planet1}  for the satellite's drag coefficient, mass, maximum and minimum drag surface areas (i.e., we use $\bar{C}_D$, $\bar{m}$, and $A_{min} \leq A \leq A_{max}$). For atmospheric density, we use a simplified version of the Harris-Priester model, as described by Montenbruck \cite{Montenbruck}, that does not consider diurnal effects. 

The relative velocity of the satellite with respect to the atmosphere is approximated based on the assumption that the atmosphere rotates with the same velocity as that of the Earth's rotation \cite{Montenbruck}:
\begin{equation} 
\vec{v}_{rel} = \vec{v}_{sat} - \vec{\omega}_{E} \times \vec{r}_{sat} \\
\label{eqn:relvel}
\end{equation}
where $\vec{v}_{sat}$ is the satellite velocity vector, $\vec{r}_{sat}$ is the satellite position vector, and $\vec{w}_{E}$ is the Earth's angular velocity about its axis.

\subsection{Simulation Model}

In this subsection we develop a simple orbital dynamical model for simulation purposes. Since two-body motion is planar in an Earth-centered inertial frame, we begin by using polar coordinates to represent the satellite orbital kinematics in the plane.
\begin{subequations}
\begin{align}
\vec{r}           &= r\underline{e}_r \\
\dot{\vec{r}}   &= \dot{r}\underline{e}_r + r\dot{\theta}\underline{e}_{\theta} \\
\ddot{\vec{r}} &= \left( \ddot{r}-r\dot{\theta}^2 \right)\underline{e}_r + \left( 2\dot{r}\dot{\theta} + r\ddot{\theta} \right)\underline{e}_{\theta}
\end{align}
\end{subequations}
We denote the magnitude of the radial position with $r$ and the angular position with $\theta$. We use $\underline{e}_r$, $\underline{e}_{\theta}$, and $\underline{e}_N$ as the unit vectors in the radial, tangential, and normal directions of the orbital plane, respectively.

If we include the right hand terms of (1), we get the following equations of motion:
\begin{subequations}
\begin{align}
\ddot{r} &= r\dot{\theta}^2 - \frac{\mu_{\scriptscriptstyle E}}{r^2} + (\vec{a}_{perturb})_r \\
\ddot{\theta} &= \frac{1}{r} \left( -2\dot{r}\dot{\theta} + (\vec{a}_{perturb})_{\theta} \right)
\end{align}
\end{subequations}

We now introduce three assumptions that regard the perturbation acceleration terms. First, we ignore all perturbations except for atmospheric drag (i.e., $\vec{a}_{perturb}$ $\approx$ $\vec{a}_{atmdrag}$). While differential drag takes advantage of the fact that atmospheric drag can secularly affect the in-plane size and shape of an orbit (i.e., its semi-major axis and eccentricity), the other dominant perturbations mainly cause secular changes that are out-of-plane, which we do not have control over. Thus, we omit those perturbations in our simplified simulation model.

Second, we assume near-circular orbits where the magnitude of the satellite's velocity vector in the tangential direction of the orbital plane is significantly larger than in the radial direction. Since atmospheric drag is antiparallel to the the velocity vector, we ignore the radial acceleration component of the drag perturbation. That is, $(\vec{a}_{atmdrag})_r$ $\approx$ $0$.

We are left with the following approximated equations of motion:
\begin{subequations}
\begin{align}
\ddot{r} &= r\dot{\theta}^2 - \frac{\mu_{\scriptscriptstyle E}}{r^2} \\
\ddot{\theta} &= \frac{1}{r} \left( -2\dot{r}\dot{\theta} + (\vec{a}_{atmdrag})_{\theta} \right)
\end{align}
\end{subequations}

Finally, we must approximate the Earth's angular velocity in the coordinate frame of the orbital plane so that we may estimate the satellite's velocity with respect to the atmosphere (\ref{eqn:relvel}). Based on the inclination ($\phi \mid 0\leq \phi \leq 180^\circ$) of a satellite's orbit, we only consider the component of the Earth's angular velocity that is about the normal axis of the orbital plane. We assume that both the angular velocity of the Earth about its axis and the inclination of the orbit are constant.
\iffalse
\begin{subequations}
\begin{align} 
\vec{v}_{rel} &= r\dot{\theta}\underline{e}_{\theta} - \left(\omega_{\scriptscriptstyle E}cos(\phi) \underline{e}_{N} \times r \underline{e}_{r} \right) \\
                    &= r\left( \dot{\theta}-\omega_{\scriptscriptstyle E}\text{cos}(\phi) \right) \underline{e}_{\theta}
\end{align}
\end{subequations}
\fi
\begin{subequations}
\begin{align} 
\vec{v}_{rel} &= r\dot{\theta}\underline{e}_{\theta} - \left(\bar{\omega}_{\scriptscriptstyle E}cos(\bar{\phi}) \underline{e}_{N} \times r \underline{e}_{r} \right) \\
                    &= r\left( \dot{\theta}-\bar{\omega}_{\scriptscriptstyle E}\text{cos}(\bar{\phi}) \right) \underline{e}_{\theta}
\end{align}
\end{subequations}

Thus, for a near-polar orbit ($\phi \approx 90^\circ$), the relative speed of the satellite is essentially the tangential speed of the satellite (i.e., the speed of the atmosphere is negligible). For a prograde, equatorial orbit ($\phi=0$), the relative speed of the satellite is slower since the satellite's motion is parallel with the velocity of the atmosphere. For a retrograde, equatorial orbit ($\phi=180^\circ$), the relative speed is faster since the satellite's motion is antiparallel with the atmospheric velocity vector.
 
To summarize, we use the following equations to simulate the approximated orbital motion of a satellite ($\omega := \dot{\theta}$):
\iffalse
\begin{subequations}
\begin{align}
\ddot{r} &= r\omega^2 - \frac{\mu_{\scriptscriptstyle E}}{r^2} \\
\ddot{\theta} &= \frac{1}{r} \left( -2\dot{r}\omega-\frac{1}{2}\frac{C_D}{m}\rho r^2\left( \omega-\omega_{\scriptscriptstyle E}\text{cos}(\phi) \right)^2 A \right)
\end{align}
\end{subequations}
\fi
\begin{subequations}
\begin{align}
\ddot{r} &= r\omega^2 - \frac{\mu_{\scriptscriptstyle E}}{r^2} \\
\ddot{\theta} &= \frac{1}{r} \left( -2\dot{r}\omega-\frac{1}{2}\frac{\bar{C}_D}{\bar{m}}\rho(r) \lvert \vec{v}_{rel}(r, \omega) \rvert^2 A \right)
\end{align}
\end{subequations}
where the notation $\rho(r)$ and $\vec{v}_{rel}(r,\omega)$ is used to denote that the atmospheric density and relative velocity terms are dependent on the satellite's radius and angular velocity.

\subsection{Approximate Discrete-Time Model}

We introduce a discrete-time dynamical model of the $i^{th}$ satellite for use in our optimization problem:
\begin{subequations}
\begin{align}
r_i(k+1) &= r_i(k) + \Delta t \cdot S^{R}(r_i(k), \omega_i(k)) \cdot u_i(k) \\
\omega_i(k+1) &= \omega_i(k) + \Delta t \cdot S^{\Omega}(r_i(k), \omega_i(k)) \cdot u_i(k) \\
%\intertext{It then follows that the angular position evolves by:}
\theta_i(k+1) &= \theta_i(k) + \Delta t \cdot \omega_i(k) \\
                      & \ \ \ \ \ \ \ \ \ + \tfrac{1}{2} \Delta t^2 \cdot S^{\Omega}(r_i(k), \omega_i(k)) \cdot u_i(k) \nonumber
\end{align}
\label{eqn:discretemodels}%
\end{subequations}
Note that our control input is the cross-sectional surface area of the satellite (i.e., $u:=A$).
\iffalse
\begin{subequations}
\begin{align}\nonumber
S^{R}(\bar{r}_i(k), \bar{\omega}_i(k)) \ \ \ &\text{same as argument name}\\ \nonumber
S^{\Omega}(\bar{r}_i(k), \bar{\omega}_i(k)) \ \ \ &\text{same as argument name}\\ \nonumber
S^{1}(\bar{r}_i(k), \bar{\omega}_i(k)) \ \ \ &\text{confuse with raise to the power}\\ \nonumber
S^{2}(\bar{r}_i(k), \bar{\omega}_i(k)) \ \ \ &\text{confuse with raise to the power}\\ \nonumber
S(\bar{r}_i(k), \bar{\omega}_i(k)) \ \ \ &\text{too many other capital letters D, U, A}\\ \nonumber
R(\bar{r}_i(k), \bar{\omega}_i(k)) \ \ \ &\text{too many other capital letters D, U, A}\\ \nonumber
T(\bar{r}_i(k), \bar{\omega}_i(k)) \ \ \ &\text{confuse with time step T}\\ \nonumber
S^{R}(\bar{r}_i(k), \bar{\omega}_i(k)) \ \ \ &\text{I like it}\\ \nonumber
S^{\Omega}(\bar{r}_i(k), \bar{\omega}_i(k)) \ \ \ &\text{I like it}\\ \nonumber
S^{\rm rad}(\bar{r}_i(k), \bar{\omega}_i(k)) \ \ \ &\text{takes too much space}\\ \nonumber
S^{\rm tang}(\bar{r}_i(k), \bar{\omega}_i(k)) \ \ \ &\text{takes too much space}\\ \nonumber
S^{\rm radial}(\bar{r}_i(k), \bar{\omega}_i(k)) \ \ \ &\text{takes too much space}\\ \nonumber
S^{\rm tangential}(\bar{r}_i(k), \bar{\omega}_i(k)) \ \ \ &\text{takes too much space}\\ \nonumber
\end{align}
\end{subequations}
\fi

The values for $S^{R} \left( \cdot \right)$ and $S^{\Omega} \left( \cdot \right)$ \iffalse$S_{r_i} \left( r_i \left(k \right),\omega_i \left(k \right) \right)$ and $S_{\omega_i}\left( r_i \left(k \right),\omega_i \left(k \right) \right)$\fi describe how the impact of the input changes depending on the current state of the satellite. For example, for any given cross-sectional surface area, a satellite will experience greater atmospheric drag force at lower altitudes than at higher altitudes. This relationship is captured using the Gaussian variation of parameters (VOP) form of the equations of motion. These equations are used to approximate the rates of change of the time-varying elements in the solution for the unperturbed, two-body system due to small perturbing forces. Vallado \cite{Vallado} shows that the average rate of change in the semi-major axis of an orbit and the angular speed of the satellite can be expressed in terms of the atmospheric drag perturbation. By applying Vallado's results to a near-circular orbit, we find the following approximate relationships:
\begin{subequations}
\begin{align}
S^{R}(r,\omega) &= -\frac{\bar{C}_D}{\bar{m}}\rho(r) \lvert \vec{v}_{rel}(r,\omega) \rvert^2 \sqrt{\frac{r^3}{\mu_{\scriptscriptstyle E}}} \\
S^{\Omega}(r,\omega) &= \frac{3}{2} \frac{\bar{C}_D}{\bar{m}}\rho(r) \lvert \vec{v}_{rel}(r,\omega) \rvert^2 \frac{1}{r}
\end{align}
\label{eqn:SrSw}
\end{subequations}

\section{Optimization Problem}

Our goal is to spread out an initial cluster of satellites in low Earth orbit so that there is equal spacing between each satellite of the shared orbital plane. We would like to complete this constellation formation maneuver in a fixed number of days while maximizing the operational lifetime of the constellation. The operational lifetime can be defined as the total number of days that all of the satellites remain in orbit. Since atmospheric density increases exponentially as the altitude decreases, a satellite under atmospheric drag experiences very rapid orbital decay as its altitude drops. Thus, our objective is to minimize the drop in altitude of the constellation, which we achieve by maximizing the altitude of the lowest satellite in the constellation at the final time step $T$ of the optimization problem:
\begin{equation}
\begin{aligned}
& \underset{U}{\text{maximize}}
& & \underset{i = 1, \ldots, N}{\text{min}}  \ \ \ \ r_i(T)
\end{aligned}
\end{equation}
Note that if we ignore oblateness and assume a spherical Earth, maximizing the altitude of a satellite is equivalent to maximizing the magnitude of its radius. The decision variables are contained in the vector $U = [u_1(0),\ldots,u_1(T-1),\ldots,u_N(0),\ldots,u_N(T-1)]^T$. So, for a constellation of $N$ satellites and a total of $T$ time steps in which the problem is feasible, there are $NT$ decision variables.

To achieve equal angular spacing of the satellites at the desired final time step $\text{T}$, we use the following inequality constraint in our optimization problem:
\begin{equation}
{\lVert \text{D} \cdot \theta(T) - \Delta_{des} \rVert}_{\infty} \leq \epsilon_{\theta} 
\end{equation}
where $\theta(T) = \begin{bmatrix} \theta_1(T), \ \theta_2(T), \ \theta_3(T), \ \hdots \ , \theta_N(T) \end{bmatrix}^\text{T}$ is the angular position state vector at time $T$. As a convention used throughout this paper, we define $\theta_1(t)$ as the angular distance traveled by an arbitrarily designated ``lead" satellite in the orbital plane. We use the same vector notation for angular velocity $\omega(T)$ and radius $r(T)$.

The matrix $D$ is defined as
\begin{equation}
\text{D} :=
\begin{bmatrix}
     1& $-$1&          &         &            \\
       &      1&   $-$1&         &            \\
       &        &\ddots&\ddots&            \\
       &        &          &        1& $-$1   \\
 $-$1&       &          &          &      1   \\
\end{bmatrix} 
\in \mathbb{R}^{N\times N}
\end{equation}

so that
\begin{equation}
\text{D} \cdot \theta(T)
 = 
\renewcommand{\arraystretch}{1.00}
\begin{bmatrix} \theta_1(T) - \theta_2(T) \\ \theta_2(T) - \theta_3(T) \\ \vdots \\ \theta_{N-1}(T) - \theta_N(T) \\ \theta_{N}(T) - \theta_1(T) \end{bmatrix} \in \mathbb{R}^{N\times1}
\end{equation}
represents the angular separation between adjacent pairs of satellites at the final time step $\text{T}$.

We define $\Delta_{des} := \begin{bmatrix} \frac{2\pi}{N},\frac{2\pi}{N}, \ \hdots \ , \frac{2\pi}{N}, \text{-}\frac{2\pi}{N}(N-1) \end{bmatrix}^\text{T}  \in \mathbb{R}^{N\times1}$ as the vector containing the desired angular spacings between each adjacent pair of satellites so that the entire constellation is equally spaced. The difference $\text{D} \cdot \theta(T)-\Delta_{des}$ results in a vector containing the angular spacing errors between adjacent pairs. By constraining the maximum of the absolute values of these errors at time $T$ to be less than or equal to some angular position error tolerance $\epsilon_{\theta}$, we may achieve approximately equal spacing.

Similarly, by constraining the angular velocities of adjacent satellites to be effectively zero, we can ensure that the constellation will tend to remain equally spaced in the future and not just at that instance:
\begin{equation}
{\lVert \text{D} \cdot \omega(T) \rVert}_{\infty} \leq \epsilon_{\omega} 
\end{equation}
where $\epsilon_{\omega}$ is an angular velocity error tolerance close to zero. 

We also impose input constraints since our control authority is limited by the actual physical dimensions of the satellite:
\begin{equation}
U_{min}  \leq U \leq U_{max}
\end{equation}

Given the initial state vectors $r(0)$, $\omega(0)$, and $\theta(0)$, we now summarize the optimization problem here:
\begin{equation}
\begin{aligned}
& \underset{U}{\text{maximize}}
& & \underset{i = 1, \ldots, N}{\text{min}}  \ \ \ \ r_i(T)\\
& \text{subject to}
& & {\lVert \text{D} \cdot \theta(T) - \Delta_{des} \rVert}_{\infty} \leq \epsilon_{\theta} \\
& & & {\lVert \text{D} \cdot \omega(T) \rVert}_{\infty} \leq \epsilon_{\omega}  \\
& & & U_{min}  \leq U \leq U_{max} \\
\end{aligned}
\end{equation}
which can be restated in the following form by introducing an extra decision variable $t$:
\begin{equation}
\begin{aligned}
& \underset{U, \; t}{\text{minimize}}
& & \ \ \ \ \ t \\
& \text{subject to}
& & \text{-}r(T) \leq t \cdot \mathbf{1}^{N \times 1} \\
& & & {\lVert \text{D} \cdot \theta(T) - \Delta_{des} \rVert}_{\infty} \leq \epsilon_{\theta} \\
& & & {\lVert \text{D} \cdot \omega(T) \rVert}_{\infty} \leq \epsilon_{\omega}  \\
& & & U_{min}  \leq U \leq U_{max}  \ . \\
\end{aligned}
\end{equation}
We note that $r(T)$, $\omega(T)$, and $\theta(T)$ do not depend linearly on the input $U$. To obtain a linear program, we first precompute reference trajectories $\bar{r_i}(\cdot)$ and $\bar{\omega_i}(\cdot)$ by using equations (\ref{eqn:discretemodels}) and (\ref{eqn:SrSw}) with the conservative assumption that each satellite is under minimum drag input until final time step $T$ (i.e., $U = U_{min}$). We then use the following relationships:
\begin{subequations} 
\begin{align}
r_i(k+1) &= r_i(k) + \Delta t \cdot S^{R} ( \bar{r_i} (k),\bar{\omega_i} (k)) \cdot u_i(k) \\
\omega_i(k+1) &= \omega_i(k) + \Delta t \cdot S^{\Omega}( \bar{r_i} (k),\bar{\omega_i} (k)) \cdot u_i(k) \\
\theta_i(k+1) &= \theta_i(k)+ \Delta t \cdot \omega_i(k) \\
                     & \ \ \ \ \ \ \ \ \ + \tfrac{1}{2} \Delta t^2 \cdot S^{\Omega} ( \bar{r_i} (k),\bar{\omega_i} (k)) \cdot u_i(k) \nonumber
\end{align}
\label{eqn:discreteapproximatemodels}%
\end{subequations}
where we have substituted the reference trajectories in $S^R(\cdot)$ and $S^\Omega(\cdot)$ so that the equations are linear but time-varying.

We estimate $r(T)$, $\omega(T)$, and $\theta(T)$ from:
\begin{subequations}
\begin{align}
r_i(T) &= r_i(0) + \Delta t \cdot \sum_{k=0}^{T-1} \left\{ S^{R} ( \bar{r_i} (k),\bar{\omega_i} (k)) \cdot u_i(k) \right\} \\
\omega_i(T) &= \omega_i(0) + \Delta t \cdot \sum_{k=0}^{T-1} \left\{ S^{\Omega} (\bar{r_i} (k),\bar{\omega_i} (k)) \cdot u_i(k) \right\} \\
\theta_i(T) &=  \theta_i(0) + \Delta t \cdot \sum_{k=0}^{T-1} \omega_i(k) \nonumber \\ 
                 & \ \ + \tfrac{1}{2} \Delta t^2 \cdot \sum_{k=0}^{T-1} \left\{ S^{\Omega} ( \bar{r_i} (k),\bar{\omega_i} (k)) \cdot u_i(k) \right\} \nonumber \\
		&=  \theta_i(0) + \Delta t T \omega_i(0) \\
		& \ \ + \Delta t^2 \cdot \sum_{k=0}^{T-1} (T-k-\tfrac{1}{2}) \left\{ S^{\Omega} ( \bar{r_i} (k),\bar{\omega_i} (k)) \cdot u_i(k) \right\} \nonumber
\end{align}
\label{eqn:linearcombinations}
\end{subequations}
which can be expressed in matrix form: 
\begin{subequations}
\begin{align}
r(T) &= r(0) + \Delta t \cdot \bar{S}^{R} \cdot U\\
\omega(T) &= \omega(0) + \Delta t \cdot \bar{S}^{\Omega} \cdot U \\
\theta(T) &=  \theta(0) + \Delta t \cdot T \cdot \omega(0) + \Delta t^2 \cdot \bar{S}^{\alpha} \cdot U
\end{align}
\end{subequations}
where $r(0)$, $\omega(0)$ and $\theta(0)$ $\in \mathbb{R}^{N\times 1}$ are the initial state vectors and $\bar{S}^R$, $\bar{S}^{\Omega}$ are large matrices of the form:

\begin{equation}
\bar{S}^R = \begin{bmatrix} \bar{S}^R_1& & \\ & \ddots & \\ & & \bar{S}^R_N \end{bmatrix}, \
\bar{S}^{\Omega} = \begin{bmatrix} \bar{S}^\Omega_1 & & \\ & \ddots & \\ & & \bar{S}^\Omega_N \end{bmatrix} \in \mathbb{R}^{N\times (N \cdot T)}
\end{equation}
consisting of the following row vectors along the diagonal for $i = 1,\ldots, N$: 
\begin{subequations}
\begin{align}
\bar{S}^R_i &= [ S^R(\bar{r_i}(0),\bar{\omega_i}(0)),  \ldots, S^R(\bar{r_i}(T-1),\bar{\omega_i}(T-1))] \\ %\in \mathbb{R}^{1 \times N} \\
\bar{S}^\Omega_i &= [ S^\Omega(\bar{r_i}(0),\bar{\omega_i}(0)),  \ldots, S^\Omega(\bar{r_i}(T-1),\bar{\omega_i}(T-1))] %\in \mathbb{R}^{1 \times N}
 \\
\intertext{The $\bar{S}^{\alpha}$ matrix is of the same form where the diagonal row vectors are found by calculating:}
\bar{S}^\alpha_i &= \left\{ (T-\tfrac{1}{2}) \cdot \mathbf{1}^{1 \times T} - [0, \dots, (T-1)] \right\} \circ \bar{S}^\Omega_i
\end{align}
\end{subequations}
where ($\circ$) is the element-wise product of the two row vectors. The $\bar{S}^R$, $\bar{S}^\Omega$, and $\bar{S}^\alpha$ matrices are precomputed prior to solving the program, \iffalse By assuming minimum drag conditions, we are underestimating the impact of the optimal inputs applied at each time step. Although this will result in a suboptimal solution, when applied to the simulation with a feedback strategy we are more likely to achieve our goal of forming an equally spaced constellation. \fi
which we can now express in standard form:

\begin{equation}
\begin{aligned}
& \underset{x}{\text{minimize}}
& & \ \ \ \ \ f^Tx \\
& \text{subject to}
& & Ax \leq b
\end{aligned}
\end{equation}
\\
where $x = [U, \ t]^T$ and $f^T = [\mathbf{0}^{1 \times (N \cdot T)}, \ 1]$. The matrices used to form the inequality constraints are: \\

\begin{equation}
\begin{aligned}
A = 
\begin{bmatrix*}[l]
\text{-} \Delta t \cdot \bar{S}^R,  & \text{-} \mathbf{1}^{N \times 1} \\ 
\phantom{\text{-}} \Delta t^2 \cdot D \cdot \bar{S}^\alpha, & \phantom{\text{-}} \mathbf{0}^{N \times 1} \\
\text{-} \Delta t^2 \cdot D \cdot \bar{S}^\alpha, & \phantom{\text{-}} \mathbf{0}^{N \times 1} \\
\phantom{\text{-}} \Delta t \cdot D \cdot \bar{S}^\Omega, & \phantom{\text{-}} \mathbf{0}^{N \times 1} \\
\text{-} \Delta t \cdot D \cdot \bar{S}^\Omega, & \phantom{\text{-}} \mathbf{0}^{N \times 1} \\
\phantom{\text{-}} \mathbb{I}^{(N \cdot T) \times (N \cdot T)}, & \phantom{\text{-}} \mathbf{0}^{(N \cdot T) \times 1} \\
\text{-} \mathbb{I}^{(N \cdot T) \times (N \cdot T)}, & \phantom{\text{-}} \mathbf{0}^{(N \cdot T) \times 1} \\ 
\end{bmatrix*} 
\end{aligned}
\end{equation}

and

\begin{equation}
\begin{aligned}
b = 
\begin{bmatrix}
r(0) \\
\epsilon_{\theta} \cdot \mathbf{1}^{N \times 1} - D \cdot \left[ \theta (0) + \Delta t \cdot T \cdot \omega(0) \right] + \Delta_{des} \\
\epsilon_{\theta} \cdot \mathbf{1}^{N \times 1} + D \cdot \left[ \theta (0) + \Delta t \cdot T \cdot \omega(0) \right] - \Delta_{des} \\
\epsilon_{\omega} \cdot \mathbf{1}^{N \times 1} - D \cdot \omega (0) \\
\epsilon_{\omega} \cdot \mathbf{1}^{N \times 1} + D \cdot \omega (0) \\
\phantom{\text{-}} u_{max} \cdot \mathbf{1}^{(N \cdot T) \times 1} \\
{\text{-}} u_{min} \cdot \mathbf{1}^{(N \cdot T) \times 1} 
\end{bmatrix} 
\end{aligned} 
\end{equation}
\\

\section{SIMULATION RESULTS}

\iffalse 
We take inspiration from the Planet Doves constellation which employs about a hundred 3U-size satellites in a sun-synchronous, low Earth orbit at an ideal altitude of 475 km. The satellites are launched from a single launch vehicle after which they are phased apart into an equally-spaced constellation. In our problem, we also assume that all satellites begin at the same angular position in a 475 km Sun-synchronous circular orbit. The launch vehicle ejects each satellite at a different orientation with respect to the orbital motion. Once fully deployed, the constellation would resemble a cluster with each satellite having a slightly different radial position, angular position and velocity with respect to a reference line in the orbital plane. \fi

The simulation results in this section are based on $N = 105$ satellites beginning at the same initial states, corresponding to a Sun-synchronous, circular orbit at an altitude of 475 km: $\theta_i(0) = 0$, $r_i(0) = r_0 $ and $\omega_{i}(0) = \sqrt{\tfrac{\mu_{\scriptscriptstyle E}}{r_i(0)^3}}$ for all $i = 1, \ldots, N$. 

In our linear program, we set the angular separation and velocity error tolerances at $\epsilon_{\theta} = 0.1$ degrees and $\epsilon_{\omega} = 1e\text{-}18$ rad/s, respectively. The solution to the program corresponds to optimal drag area commands (i.e., desired cross-sectional surface areas) that are sent simultaneously to all satellites once every 24 hours. The drag area commands are allowed to take on any value within a continuous range between the minimum and maximum possible surface areas. 

\subsection{Open-loop versus Feedback}

\begin{figure}[h]
\centering
\includegraphics[width=.5\textwidth]{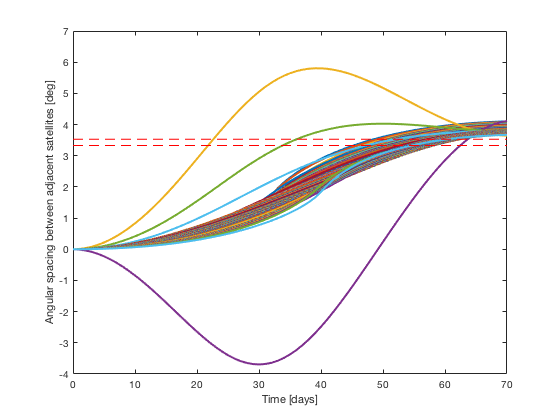}
\caption{Optimal input commands applied in open-loop result in final angular spacing values \iffalse($\theta_1 - \theta_2$, $\ldots$, $\theta_{N-1} - \theta_N$) \fi that fail to land within the designed error tolerance thresholds (represented by dashed lines). Note that the angular spacing between the $N^{th}$ and $1^{st}$ satellites ($\theta_N - \theta_1$), which should converge to -$\frac{2\pi}{N}(N-1)$, is not included in this figure.}
\label{fig:DeltasOpenLoop}
\end{figure}

Starting at the initial conditions, we determine that the linear program is feasible with a horizon of $T = 71 $ days. We apply the optimal input commands in open-loop and find that after 71 days, the angular spacings between adjacent satellites tend to converge towards the desired value of $\frac{2\pi}{N}$ (see  Fig. \ref{fig:DeltasOpenLoop}). However, due to the approximations made in the linear model (\ref{eqn:discreteapproximatemodels}), none of the angular spacings satisfy the angular spacing error tolerance that we defined. Furthermore, although the program expects a maximum altitude drop of 10.79 km at the end of 71 days, the open-loop simulation results in a worse 11.64 km altitude drop.

\begin{figure}[h]
\centering
\includegraphics[width=.5\textwidth]{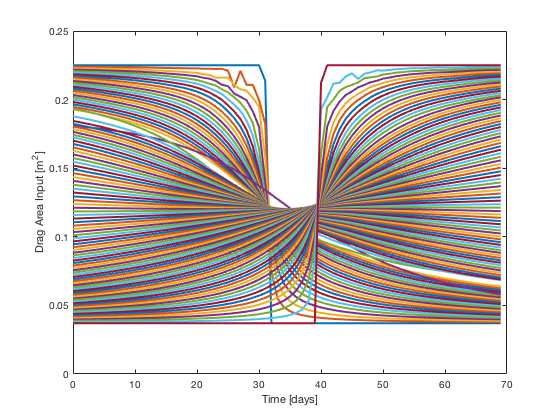}
\caption{Optimal input commands for all $N$ satellites applied in open-loop each day until horizon $T = 71$ days.}
\label{fig:InputsOpenLoop}
\end{figure}

Fig. \ref{fig:InputsOpenLoop} shows how the solution to our initial optimization problem is to utilize the whole range of input values, from minimum to maximum, and to give each satellite a different input value  at time step $k=0$ based on its arbitrary numbering (i.e., the $1^{st}$ satellite receives a maximum drag command while the $N^{th}$ satellite receives a minimum drag command). By the final time step T, the input value for any particular satellite is  ``flipped" in intensity compared to its initial value. For example, the input for the $1^{st}$ satellite changes from an initial maximum drag command to a final minimum drag command. All the while, the ``middle" satellite receives a relatively steady input value at all steps. The result is that the satellites which are arbitrarily assigned low or high numbered positions (i.e., $1$, $2$, $3$, ... or  ... $(N-2)$, $(N-1)$, $N$) experience larger changes in input than the satellites placed towards the middle of the constellation position numbering scheme.

Due to the dissatisfying open-loop performance, we then use a model predictive control (MPC) approach where the linear program is solved at the beginning of each time step but only the first set of control inputs in the sequence is applied to the satellites. At the first time step, we solve for the horizon $T$. At each subsequent time step, the problem is reformulated and solved again but with a shrinking horizon (i.e., at $k = 0$, horizon $= T$, at $k = 1$, horizon $= T-1$). By solving the program at each time step with updated states, we are able to correct for prediction errors and the effect of un-modeled perturbations.

\begin{figure}[h]
\centering
\includegraphics[width=.5\textwidth]{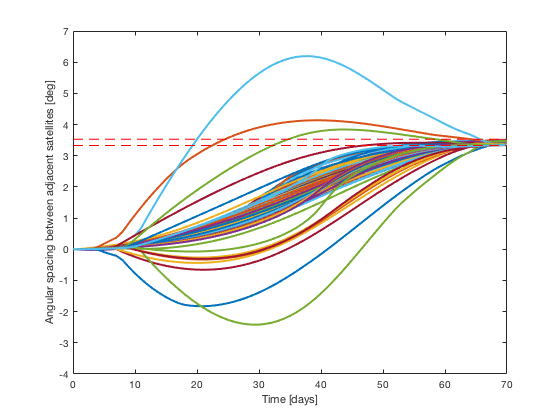}
\caption{Angular spacing between adjacent pairs of satellites change from $0^{\circ}$ to $\tfrac{360^{\circ}}{N} \pm \epsilon_{\theta}$. Note that the angular spacing between the $N^{th}$ and $1^{st}$ satellites ($\theta_N - \theta_1$), which should converge to -$\frac{2\pi}{N}(N-1)$, is also not included in this figure.}
\label{fig:spacinghorizon}
\end{figure}

\begin{figure}[h]
\centering
\includegraphics[width=.5\textwidth]{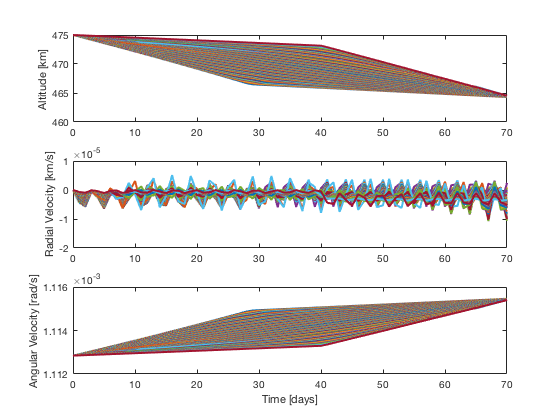}
\caption{At the horizon $T$, the altitudes (and angular velocities) of the satellites converge, ensuring that the constellation will remain in the desired configuration with minimal control effort for the remainder of the constellation lifetime.}
\label{fig:stateshorizon}
\end{figure}

\begin{figure}[h]
\centering
\includegraphics[width=.5\textwidth]{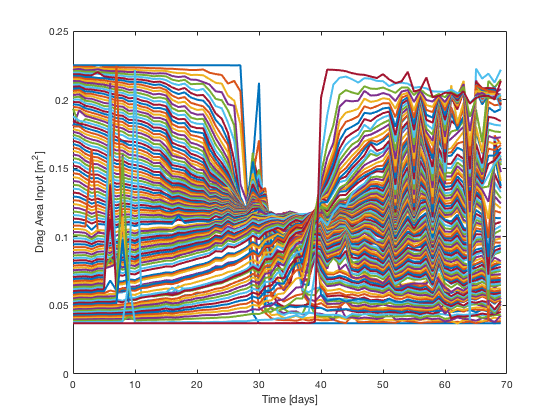}
\caption{The average level of actuation is high both in the beginning and towards the end of optimal constellation separation phase.}
\label{fig:inputshorizon}
\end{figure}

As can be seen in Fig. \ref{fig:spacinghorizon}, when the optimal inputs are applied with feedback, an equally-spaced constellation is formed within 71 days where all the angular spacings satisfy the designed angular separation error tolerance. Also, compared to the 11.64 km altitude loss in the open-loop simulation, Fig. \ref{fig:stateshorizon} shows that all the satellites under feedback control converge in altitude and angular velocity at the expense of losing only 10.71 km in altitude (i.e., we conserve 1 km of altitude), which is very close to the 10.79 km predicted by the program at time step $k=0$. This 10.71 km drop in altitude compares to a 2.84 km drop under constant minimum drag and a 19.88 km drop under maximum drag, for the same number of days. Fig. \ref{fig:stateshorizon} also shows how the altitude of each satellite is varied over time, resulting in an inversely proportional change in angular velocity. This difference in angular velocity between pairs of satellites allows the controller to adjust the angular spacings to the desired values. Fig. \ref{fig:inputshorizon} shows the input area commands that are computed and applied at the beginning of each day. 

\subsection{Trading Constellation Acquisition Time for Lifetime}

\begin{figure}[h]
\centering
\includegraphics[width=.5\textwidth]{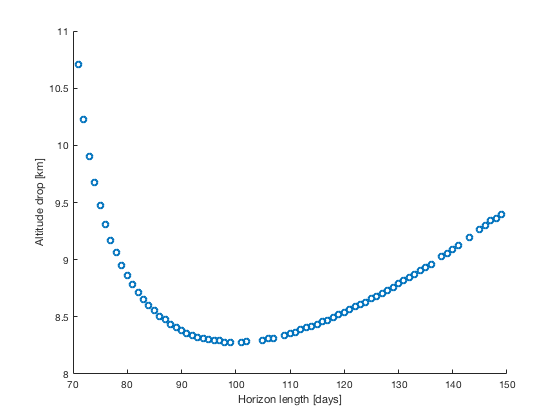}
\caption{Smallest altitude drop of 8.28 km is achieved when horizon length is 98 days.}
\label{fig: varyinghorizonlength}
\end{figure}

We increase the horizon length $T$ (i.e., number of days for the satellites to converge to the desired constellation) to determine the effect on the overall lifetime of the constellation. Fig. \ref{fig: varyinghorizonlength} shows that as we increase the horizon length from 71 days to 98 days, we can reduce the altitude drop by 2.43 km, effectively extending the lifetime of the constellation. However, increasing the horizon beyond 98 days does not result in further improvement.

\subsection{Maintaining Constellation throughout Lifetime}

\begin{figure}[h]
\centering
\includegraphics[width=.5\textwidth]{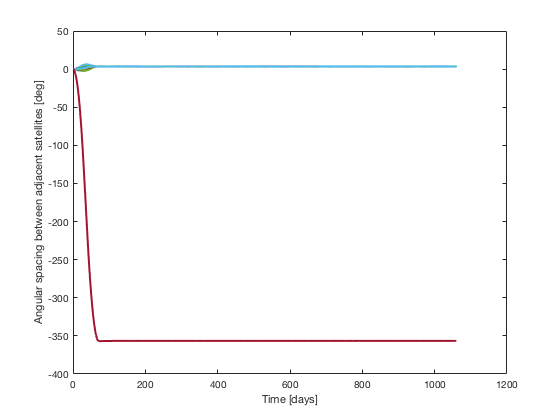}
\caption{Angular separation is maintained for the lifetime of constellation. Note that the angular difference between the $N^{th}$ and $1^{st}$ satellite reaches a value of -$\frac{360^{\circ}}{N}(N-1) \pm \epsilon_{\theta}$ .}
\label{fig:spacinglifetime}
\end{figure}

\begin{figure}[h]
\centering
\includegraphics[width=.5\textwidth]{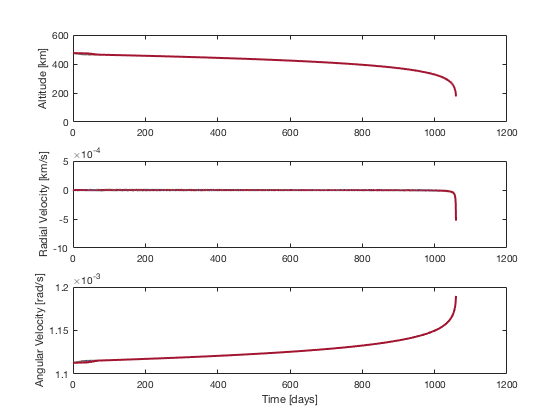}
\caption{Once equally spaced, the orbital motion of the constellation is relatively ``smooth" compared to the initial, optimal constellation separation phase.}
\label{fig:stateslifetime}
\end{figure}

\begin{figure}[h]
\centering
\includegraphics[width=.5\textwidth]{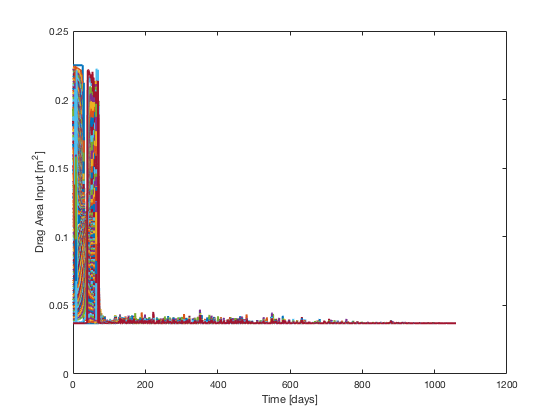}
\caption{Once equally spaced, relatively minimal actuation is required to maintain the angular separation for the duration of the constellation operational lifetime.}
\label{fig:inputslifetime}
\end{figure}

Once the equally-spaced constellation is achieved, the initial acquisition phase is complete and the satellites enter the operational mode where they are allowed to ``drift" in a minimum-drag attitude configuration. However, when the spacing between any adjacent pairs reaches above a certain threshold, we apply the optimal control strategy again, albeit with a much shorter horizon. In Fig. \ref{fig:spacinglifetime} we show that our approach is successful in maintaining the angular separations throughout the operational phase. We also observe that the orbital motion of the constellation is relatively ``smooth" in the operational phase compared to the initial acquisition phase (see Fig. \ref{fig:stateslifetime}). Furthermore, in Fig. \ref{fig:inputslifetime} we see that the level of actuation required to maintain the constellation is significantly less than that required in the acquisition phase. 

We arbitrarily define the constellation operational lifetime to be the total number of days that none of the satellites drops to an altitude of 200 km or less, where spacecraft orbits decay rapidly. The total constellation lifetime is 1,059 days (2.90 years) under this optimal angular separation control strategy. As a comparison, a satellite under constant minimum or maximum drag would have a lifetime of approximately 1,410 days (3.86 years) or 232 days, respectively. 

\section{CONCLUSIONS}

Although the orbital dynamics are nonlinear, we found that both the altitude and angular velocity of a satellite controlled by differential drag can be approximated as linear over the relatively small operating range of a singe day. Thus, the solution from our linear program, even when applied in open-loop, provides reasonable performance in forming the equally spaced constellation. To compensate for the prediction error caused by model-process mismatch and improve controller performance, we leverage the feedback mechanism provided by the shrinking-horizon MPC approach. With feedback, we are able to achieve an equally-spaced constellation that satisfies design tolerances and avoids unnecessary control action that reduces operational lifetime. We also observed that we can increase the operational lifetime of the constellation by allowing it to form over a longer time frame. However, the tradeoff can only be made until a certain point at which increasing the horizon further results in decreased lifetime. Finally, we show that the constellation can be maintained throughout its lifetime by applying the same optimal control strategy when the angular spacing errors drift above a designed threshold value.
\iffalse
For immediate future work, we can use more sophisticated models and consider other, more practical constraints:
\begin{itemize}
\item Include other perturbations in the simulation model, such as solar radiation pressure and oblate Earth gravity, to further analyze controller robustness.
\item Use a more accurate atmospheric density model that takes into account the day-night variation in density.
\item In practice, not all the satellites receive the input commands at the same time due to ground station availability. At worst, the time between sending input commands to the $1^{st}$ and to the $N^{th}$ satellite could be approximately 90 minutes delayed. We will consider improvements to our linear program that will take into account the effect of such communication latency.
\item Consider energy and power constraints caused by the use of reaction wheels (i.e., to maintain the desired attitude for conducting differential drag maneuvers) as well as the onboard battery storage capacity and solar energy conversion capability of typical cubesats.
\end{itemize}

This work can also be extended to acquire constellations where satellites may be required to enter tight formations and/or perform rendezvous maneuvers. Furthermore, we may also consider the use of nontraditional actuators (e.g., solar sails, electric propulsion) for applications in interplanetary or deep space missions.
 \fi
 
\addtolength{\textheight}{-12cm}   % This command serves to balance the column lengths

%%%%%%%%%%%%%%%%%%%%%%%%%%%%%%%%%%%%%%%%%%%%%%%%%%%%%%%%%%%%%%%%%%%%%%%%%%%%%%%%

%%%%%%%%%%%%%%%%%%%%%%%%%%%%%%%%%%%%%%%%%%%%%%%%%%%%%%%%%%%%%%%%%%%%%%%%%%%%%%%%

\iffalse
%%%%%%%%%%%%%%%%%%%%%%%%%%%%%%%%%%%%%%%%%%%%%%%%%%%%%%%%%%%%%%%%%%%%%%%%%%%%%%%%
\section*{APPENDIX}

Appendixes should appear before the acknowledgment.
\fi

\section*{ACKNOWLEDGMENT}

Thanks to Professor Simone D'Amico of the Department of Aeronautics and Astronautics at Stanford University for introducing this problem in the AA279A:$\it{\ Space \ Mechanics}$ course of Winter 2017. The authors gratefully acknowledge support from the National Science Foundation under grant ECCS-1405413.  Andrew Packard acknowledges the generous support from the FANUC Corporation.\\

%%%%%%%%%%%%%%%%%%%%%%%%%%%%%%%%%%%%%%%%%%%%%%%%%%%%%%%%%%%%%%%%%%%%%%%%%%%%%%%%
\iffalse
References are important to the reader; therefore, each citation must be complete and correct. If at all possible, references should be commonly available publications.
\fi

\end{document}